\documentstyle[psfig]{mn}

\newif\ifAMStwofonts
\AMStwofontstrue

\ifoldfss
  \newcommand{\rmn}[1] {{\rm #1}}

  \ifCUPmtlplainloaded \else
    \NewTextAlphabet{textbfit} {cmbxti10} {}
    \NewTextAlphabet{textbfss} {cmssbx10} {}
    \NewMathAlphabet{mathbfit} {cmbxti10} {} 
    \NewMathAlphabet{mathbfss} {cmssbx10} {} 
  \fi
  \ifAMStwofonts
    \ifCUPmtlplainloaded \else
      \NewSymbolFont{upmath} {eurm10}
      \NewSymbolFont{AMSa} {msam10}
      \NewMathSymbol{\upi}     {0}{upmath}{19}
      \NewMathSymbol{\umu}     {0}{upmath}{16}
      \NewMathSymbol{\upartial}{0}{upmath}{40}
      \NewMathSymbol{\leqslant}{3}{AMSa}{36}
      \NewMathSymbol{\geqslant}{3}{AMSa}{3E}

    \fi
  \fi
\fi 

\ifnfssone
  \newmathalphabet{\mathit}
  \addtoversion{normal}{\mathit}{cmr}{m}{it}
  \addtoversion{bold}{\mathit}{cmr}{bx}{it}
  \newcommand{\rmn}[1] {\mathrm{#1}}

  \newmathalphabet{\mathbfit} 
  \addtoversion{normal}{\mathbfit}{cmr}{bx}{it}
  \addtoversion{bold}{\mathbfit}{cmr}{bx}{it}
  \newmathalphabet{\mathbfss} 
  \addtoversion{normal}{\mathbfss}{cmss}{bx}{n}
  \addtoversion{bold}{\mathbfss}{cmss}{bx}{n}
  \ifAMStwofonts
    \ifCUPmtlplainloaded \else
      \UseAMStwoboldmath
      \makeatletter
      \new@mathgroup\upmath@group
      \define@mathgroup\mv@normal\upmath@group{eur}{m}{n}
      \define@mathgroup\mv@bold\upmath@group{eur}{b}{n}
      \edef\UPM{\hexnumber\upmath@group}
      \new@mathgroup\amsa@group
      \define@mathgroup\mv@normal\amsa@group{msa}{m}{n}
      \define@mathgroup\mv@bold\amsa@group{msa}{m}{n}
      \edef\AMSa{\hexnumber\amsa@group}
      \makeatother
      \mathchardef\upi="0\UPM19
      \mathchardef\umu="0\UPM16
      \mathchardef\upartial="0\UPM40
      \mathchardef\leqslant="3\AMSa36
      \mathchardef\geqslant="3\AMSa3E
    \fi
  \fi
\fi 

\ifnfsstwo
  \newcommand{\rmn}[1] {\mathrm{#1}}

  \DeclareMathAlphabet{\mathbfit}{OT1}{cmr}{bx}{it}
  \SetMathAlphabet\mathbfit{bold}{OT1}{cmr}{bx}{it}
  \DeclareMathAlphabet{\mathbfss}{OT1}{cmss}{bx}{n}
  \SetMathAlphabet\mathbfss{bold}{OT1}{cmss}{bx}{n}
  \ifAMStwofonts
    \ifCUPmtlplainloaded \else
      \DeclareSymbolFont{UPM}{U}{eur}{m}{n}
      \SetSymbolFont{UPM}{bold}{U}{eur}{b}{n}
      \DeclareSymbolFont{AMSa}{U}{msa}{m}{n}
      \DeclareMathSymbol{\upi}{0}{UPM}{"19}
      \DeclareMathSymbol{\umu}{0}{UPM}{"16}
      \DeclareMathSymbol{\upartial}{0}{UPM}{"40}
      \DeclareMathSymbol{\leqslant}{3}{AMSa}{"36}
      \DeclareMathSymbol{\geqslant}{3}{AMSa}{"3E}
    \fi
  \fi
\fi 

\ifCUPmtlplainloaded \else
  \ifAMStwofonts \else 
    \def\upi{\pi}
    \def\umu{\mu}
    \def\upartial{\partial}
  \fi
\fi

\title{Superhumps in V348 Pup}
\author[D. J. Rolfe et al.]
       {Daniel J. Rolfe,$^1$ Carole A. Haswell,$^1$ and Joseph Patterson$^2$\\
$^1$Department of Physics and Astronomy, The Open University, Walton Hall, Milton Keynes MK7 6AA\\
$^2$Department of Astronomy, Columbia University, 538 W. 120th St.,
 New York, New York 10027}
\date{Accepted March 2000}

\pagerange{\pageref{firstpage}--\pageref{lastpage}}
\pubyear{2000}

\begin{document}

\maketitle

\label{firstpage}

\begin{abstract}
The eclipsing novalike cataclysmic variable star V348 Pup exhibits a
persistent luminosity modulation with a period 6 per cent longer than
its 2.44 hour orbital-period ($P_{\rmn{orb}}$). This has been
interpreted as a ``positive superhump'' resulting from a slowly
precessing non-axisymmetric accretion disc gravitationally interacting
with the secondary. We find a clear modulation of mid-eclipse times on
the superhump period, which agrees well with the predictions of a
simple precessing eccentric disc model. Our modelling shows that the
disc light centre is on the far side of the disc from the donor star
when the superhump reaches maximum light. This phasing suggests a link
between superhumps in V348 Pup and late superhumps in SU UMa
systems. Modelling of the full lightcurve and maximum entropy eclipse
mapping both show that the disc emission is concentrated closer to the
white dwarf at superhump maximum than at superhump minimum. We detect
additional signals consistent with the beat periods between the
implied disc precession period and both $\frac{1}{2}P_{\rmn{orb}}$ and
$\frac{1}{3}P_{\rmn{orb}}$.
\end{abstract}

\begin{keywords}
accretion discs -- binaries: close -- binaries: eclipsing - stars:
individual: V348 Pup -- stars: cataclysmic variables
\end{keywords}

\section{Introduction}

V348 Pup (1H 0709--360, Pup 1) is a novalike cataclysmic variable
(CV): a system with a high mass transfer rate which maintains its
accretion disc in the hot, ionized, high viscosity state reached by
dwarf novae in outburst. It exhibits deep eclipses in its optical and
infrared lightcurves (Tuohy et al. 1990): it is a high inclination
system with orbital period $P_{\rmn{orb}}=2.44$ hours (Baptista et
al. 1996).

\subsection{Superhumps}

Modulations in luminosity with a period a few per cent longer than the
orbital period have been observed in many short period CVs (see
reviews in Molnar \& Kobulnicky 1992, Warner 1995, Patterson
1998a). These modulations typically take the form of a distinct
increase in luminosity, or superhump. The standard explanation of this
phenonemon is that the system contains an eccentric precessing
accretion disc.

If the accretion disc extends out far enough, the outermost orbits of
disc matter can resonate with the tidal influence of the secondary
star as it orbits the system. A 3:1 resonance can occur which results
in the disc becoming distorted to form an eccentric non-axisymmetric
shape. The tidal forces acting on this eccentric disc will cause it to
precess slowly in a prograde direction.

The superhump period, $P_{\rmn{sh}}$, is then the
 beat period between the disc precession period, $P_{\rmn{prec}}$, and 
the orbital period, $P_{\rmn{orb}}$ (Osaki 1996):
\[
\frac{1}{P_{\rmn{sh}}} = \frac{1}{P_{\rmn{orb}}} - \frac{1}{P_{\rmn{prec}}}.
\]
$P_{\rmn{sh}}$ is the period on which the relative orientation of the
line of centres of the two stars and the eccentric disc
repeats. Possible models for the light modulation on $P_{\rmn{sh}}$
are described below. This paper considers these models in relation to
our observations.

In the tidal model the superhump is a result of tidal stresses acting
on the precessing eccentric disc (Whitehurst 1998b). The light may be
due to a perturbation of the velocity field in the outer disc, leading
to azimuthal velocity gradients and crossing or converging particle
trajectories. Thus extra dissipation modulated on the superhump period
arises when the secondary sweeps past the eccentric disc. In addition,
the superhump-modulated tidal stress would lead to a
superhump-modulated angular momentum loss from the disc which would
facilitate a variation in the mass transfer rate through the disc, and
hence a modulation in disc luminosity.

The bright spot model arises from noting that the energy gained by
material in the accretion stream will depend on how far it falls
before impacting on the disc (Vogt 1981). The energy dissipated at
impact will be modulated on the superhump period since the
non-axisymmetric disc radius causes a stream-disc impact region at
varying depths in the white dwarf potential well.

Recent SPH simulations of accretions discs in AM CVn stars lead to a
third, more realistic, model in which the disc shape changes from
nearly circular to highly eccentric over the course of a superhump
period (Simpson \& Wood 1998). Superhumps arise from viscous energy
production as the distorting disc is tidally stressed. Other SPH
simulations (e.g. Murray 1996, 1998) also reveal a disc whose shape
changes, with Murray (1996) predicting superhump modulations from both
the periodic compression of the eccentric disc and the varying depth
in the primary Roche potential at which the stream impacts the disc.

Dwarf novae in super-outburst exhibit two distinct positive superhump
phenomena (Vogt 1983, Schoembs 1986). Normal superhumps appear early
in the super-outburst and fade away towards the end of the outburst
plateau to be replaced with `late' superhumps which persist into
quiescence. These late superhumps are roughly anti-phased with the
normal superhumps, and are more likely to be analogous to the
persistent superhumps seen in novalikes (Patterson 1998b), where the
system has had sufficient time to settle into a steady state. Our
extensive photometry (Section \ref{observations}) reveals similarities
between superhumps in V348 Pup and late superhumps in dwarf novae.

In Section \ref{periodsearch} we present power spectra revealing the
superhump period and additional signals close to orbital period
harmonics. In Section \ref{qi} we estimate the orbital parameters, $q$
and $i$, for V348 Pup using the average orbital lightcurve and the
superhump period. The waveform of the superhump modulation is
discussed in Section \ref{superhumpmod}. Section \ref{averages}
considers average orbital lightcurves grouped according to superhump
phase. In Section \ref{ecpar} we fit our lightcurves with a precessing
eccentric disc model, hence deducing the location of light centre of
the disc. We consider the results of maximum entropy eclipse mapping
in section \ref{eclipsemapping}. Our results and their implications
are discussed in Section
\ref{discussion}.

\section{The observations}
\label{observations}

\begin{figure}
\psfig{file=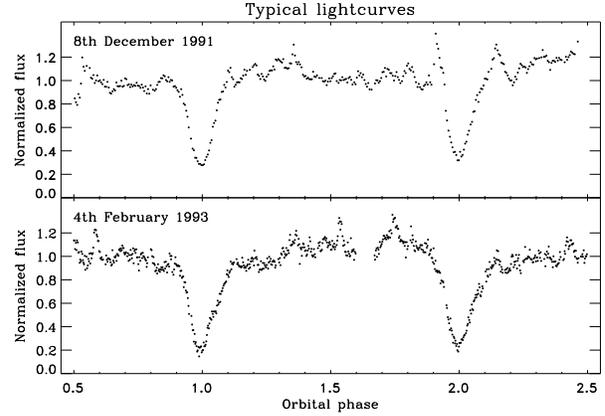,width=\columnwidth}
\caption{4 cycles of typical data from 1991 and 1993.}
\label{exampleplots}
\end{figure}

The observing campaign comprises 24 nights of rapid photometry from
December 1991, February 1993 and January 1995 (see Table
\ref{obslog}). The 1991 and 1993 observations (12 and 8 nights
respectively) were taken using the the 40-inch telescope at CTIO with
a blue copper sulphate filter. The January 1995 run consists of 4
nights of R band data. All the data have been corrected for
atmospheric extinction and the 1995 data have also been calibrated to
give an absolute flux. In 1995 the average out of eclipse R magnitude
is 15.5 mag; at mid-eclipse $R=16.8$. Examples of typical data are plotted
in Figure \ref{exampleplots}.

\begin{table}
\caption{Log of observations}
\label{obslog}
\begin{tabular}{ccccc}
Date		&   HJD start 		&Duration &Integration&	Filter	\\
		&	$-$2440000		&(hours)		 &time (sec)&\\
\hline
1991 Dec 1	&	8591.6779	&3.96		&42	&CuSO$_4$\\	
1991 Dec 2	&	8592.6138       &3.55		&42	&CuSO$_4$\\
1991 Dec 3	&	8593.6092       &5.68		&42	&CuSO$_4$\\
1991 Dec 4	&	8594.6249       &5.14		&42	&CuSO$_4$\\
1991 Dec 5	&	8595.6203       &5.37		&42	&CuSO$_4$\\
1991 Dec 6	&	8596.6328       &5.18		&42	&CuSO$_4$\\
1991 Dec 7	&	8597.6293       &5.08		&42	&CuSO$_4$\\
1991 Dec 8	&	8598.6107       &5.59		&42	&CuSO$_4$\\
\hline
1993 Feb 2	&	9020.5672       &5.56		&10	&CuSO$_4$\\
1993 Feb 3	&	9021.5549       &6.24		&10	&CuSO$_4$\\
1993 Feb 4	&	9022.5497       &6.17		&10	&CuSO$_4$\\
1993 Feb 9	&	9027.5565       &3.71		&10	&CuSO$_4$\\
1993 Feb 11	&	9029.7051       &2.36		&10	&CuSO$_4$\\
1993 Feb 12	&	9030.6851       &2.84		&10	&CuSO$_4$\\
1993 Feb 13	&	9031.5446       &6.02		&10	&CuSO$_4$\\
1993 Feb 14	&	9032.5427       &6.23		&10	&CuSO$_4$\\
1993 Feb 21	&	9039.6498       &3.20		&10	&CuSO$_4$\\
1993 Feb 22	&	9040.5572       &3.13		&10	&CuSO$_4$\\
1993 Feb 23	&	9041.5479       &2.52		&10	&CuSO$_4$\\
\hline
1995 Jan 3	&	9720.5689       &4.83		&10	&R band\\
1995 Jan 4	&	9721.5521       &3.32		&10	&R band\\
1995 Jan 5	&	9722.5509       &7.15		&10	&R band\\
1995 Jan 6	&	9723.5472       &7.24		&10	&R band\\
\hline
\end{tabular}
\end{table}

\section{Analysis}

We determined mid-eclipse timings as described in Section \ref{ecpar}
from which we determined an orbital ephemeris for this dataset
$$\rmn{T_{mid}~=~HJD~2448591.667969(85)~+~0.101838931(14)E.}$$ This is
consistent with the Baptista et al. (1996) ephemeris within the quoted
error limits. We adopt our ephemeris for this analysis. The eclipse
timings are given in Table \ref{eclipses}.

\begin{table}
\caption{Eclipse timings}
\label{eclipses}
\begin{tabular}{cccc}
HJD mid-eclipse & Error    & HJD mid-eclipse & Error    \\
$-$2440000        & estimate & -2440000        & estimate \\
\hline
\multicolumn{2}{c}{December 1991} & \multicolumn{2}{c}{February 1993} \\
   8591.769462 & 0.000121 &  9020.613001 & 0.000036 \\
   8592.686379 & 0.000096 &  9020.715111 & 0.000043 \\
   8593.705382 & 0.000120 &  9021.632451 & 0.000090 \\
   8593.806642 & 0.000017 &  9021.734218 & 0.000082 \\
   8594.722715 & 0.000090 &  9022.650034 & 0.000232 \\
   8594.824510 & 0.000162 &  9022.752227 & 0.000175 \\
   8595.640158 & 0.000104 &  9027.640565 & 0.000417 \\
   8595.742287 & 0.000201 &  9029.779356 & 0.000091 \\
   8596.657488 & 0.000082 &  9030.695522 & 0.000026 \\
   8596.759320 & 0.000100 &  9030.797304 & 0.000213 \\
   8597.676629 & 0.000068 &  9031.612099 & 0.000105 \\
   8597.778677 & 0.000083 &  9031.714122 & 0.000104 \\
   8598.694383 & 0.000058 &  9032.630438 & 0.000045 \\
   8598.796652 & 0.000090 &  9032.732446 & 0.000022 \\
\multicolumn{2}{c}{January 1995} &    9039.657734 & 0.000007 \\
   9720.653953 & 0.000003 &  9039.759941 & 0.000159 \\
   9720.756257 & 0.000300 &  9040.573722 & 0.000065 \\
   9721.571129 & 0.000069 &  9040.675596 & 0.000025 \\
   9721.672783 & 0.000132 &  9041.592531 & 0.000082 \\
   9722.589171 & 0.000043 & & \\
   9722.690736 & 0.000053 & & \\
   9723.607919 & 0.000087 & & \\
   9723.709706 & 0.000135 & & \\
   9723.811139 & 0.000068 & & \\
\hline
\end{tabular}
\end{table}

\subsection{The superhump period}
\label{periodsearch}

Before performing any analysis, we normalized each night of data by
dividing by the average out of eclipse value.

\begin{figure}
\psfig{file=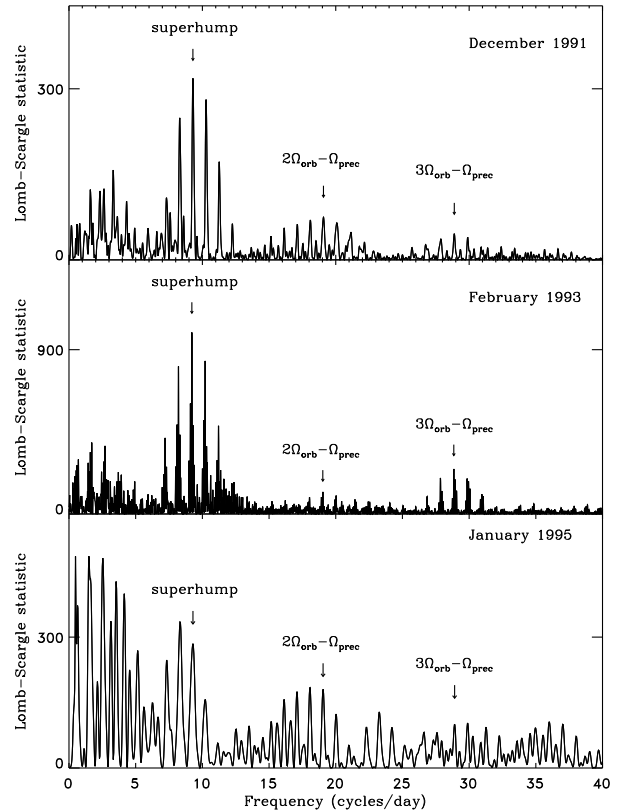,width=\columnwidth}
\caption{Power spectra. Arrows from left to right indicate the
superhump period and the sidebands of the first and second harmonics
 of the orbital period (see Table \ref{harmonics}).}
\label{fullscargle}
\end{figure}

\begin{figure}
\psfig{file=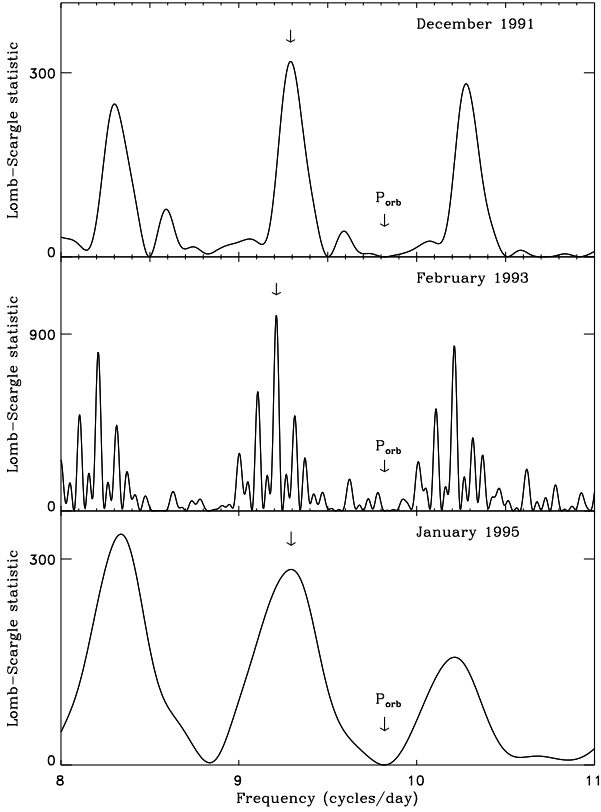,width=\columnwidth}
\caption{Detail of the power spectra shown in Figure \ref{fullscargle}.
Arrows indicate the superhump period. The orbital period
$P_{\rmn{orb}}$ is also marked, from which it is clear that the
orbital variation has been successfully removed.}
\label{scargle}
\end{figure}

To make detection of non-orbital modulations in the data easier, the
average orbital lightcurves\footnote{Superhump phase grouped average
lightcurves are shown in Figure \ref{orbcurves}.}  from each years'
observations were calculated and subtracted from the corresponding
data. The resulting lightcurves contain no orbital
variations. Lomb-Scargle periodograms were calculated for each year's
data, and are shown in Figures \ref{fullscargle} and
\ref{scargle}. The 1991 periodogram reveals a periodicity with period
0.10763~days and simple aliasing structure. The 1993 periodogram has
higher resolution, more complicated alias structure, and the strongest
periodicity at 0.10857~days. The 1995 power spectrum (having the least
time coverage) is lower resolution, however a clear signal at
0.10760~days and its aliases is present.  By comparison with the
clearer 1991 and 1993 spectra we surmise the 0.10760-day peak is the
true signal.  These periods are all close to 6 per cent greater than
$P_{\rmn{orb}}$ and correspond to the period excesses, $\epsilon$,
shown in Table \ref{shper}, defined using superhump period,
$P_{\rmn{sh}} = (1 + \epsilon) P_{\rmn{orb}}$. The inferred disc
precession periods are also shown in Table \ref{shper}.

It is notoriously difficult to determine errors in periods measured
from periodograms. To estimate the errors in the periods detected
here, various fake datasets were generated. The lightcurves were
smoothed and the residuals used to characterize the variance of the
random noise. Various types of noise with the same variance were added
to both the smoothed lightcurves and smoothed superhump modulations.
The variance in the period value determined from these different
datasets provides a measure of the error in the period measurement.
Since the smoothed data will still contain noise artefacts these
are probably underestimates. The 1993 dataset with its 20-day time
base should provide the most precise period. Therefore, the
possibility that the real 1991 and 1995 superhump periods are in fact
closer to the 1993 value than our measurements suggest should not be
ruled out, as Figure \ref{scargle} shows. However, variability in
the detected periodicity has also been seen before for superhumps
(Patterson 1998b).

Superhumping systems show a clear correlation between $\epsilon$ and
$P_{\rmn{orb}}$, with $\epsilon$ increasing with $P_{\rmn{orb}}$; the
superhump periods we detect in V348 Pup are consistent with this trend
(Figure 3, Patterson 1998b). 
 
\begin{figure}
\psfig{file=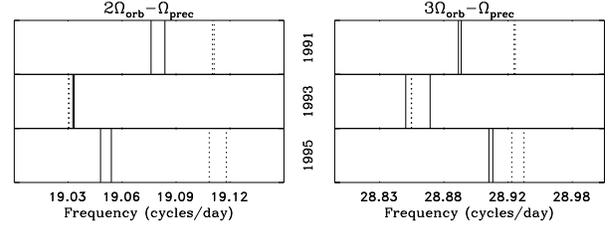,width=\columnwidth}
\caption{Sidebands of the orbital period harmonics. 
Dotted lines mark predicted frequency range for each signal, solid
lines show the measured range.}
\label{sidebands}
\end{figure}

The power spectra also reveal signals at frequencies corresponding to
sidebands of harmonics of the orbital period. The strongest such
detections are at periods corresponding to
$2\Omega_{orb}-\Omega_{prec}$ and $3\Omega_{orb}-\Omega_{prec}$
(marked with arrows in Figure \ref{fullscargle}): the predicted and
the directly measured values of these sidebands are shown in Table
\ref{harmonics}, and graphically in Figure \ref{sidebands}.
The errors in these periods were estimated using the method described
earlier.  The detected periods do not all agree to within the
estimated errors, suggesting that the error estimates may be a little
too low, as expected; the highest quality 1993 data agrees best.

The simplest way to produce these sidebands is by modulating the
brightness or visibility of the superhump with orbital phase. If we
consider the orbital lightcurve as a sum of Fourier components with
frequencies $n\Omega_{orb}$, then following the approach of Warner
(1986) and Norton, Beardmore \& Taylor (1996), the eclipse of the
superhump light source will produce signals at frequencies
$(n+1)\Omega_{orb}-\Omega_{prec}$. Signals at frequencies
$(n-1)\Omega_{orb}+\Omega_{prec}$ are also predicted but we find no
evidence of these; perhaps they are nullified by other signals of the
same frequency in antiphase. In Section
\ref{averages} we present evidence for a correlation between superhump
amplitude and orbital phase; this modulation of superhump amplitude
with orbital phase could also lead to the observed sideband signals.
The SPH models of Simpson \& Wood (1998) predict the formation of
double armed spiral density waves in the disc whose rotation rate,
they suggest, might lead to observed signals at about three times the
superhump frequency. They also suggest that viewing these structures
from non-zero inclination could lead to the detection of further
frequencies, although they do not make precise
predictions. Observations of the dwarf nova IP Peg in outburst have
revealed evidence of such spiral structure (Steeghs, Harlaftis \&
Horne 1997).

There is no significant signal around period $P_{prec}=$1.6 -- 1.9
days in either the normalized or un-normalized lightcurves. The disc
precession period is similarly absent in other persistently
superhumping systems, notably AM CVn where $P_{prec}$ is clearly
revealed by absorption line spectroscopy (Patterson, Halpern and
Shambrook 1993).

\begin{table}
\caption{Measured superhump characteristics, period $P_{sh}$, the period excesses $\epsilon$,
the inferred disc precession periods $P_{prec}$ and the consequent
estimates of the mass ratio $q$.}
\label{shper}
\begin{tabular}{ccccccc}
	&$P_{\rmn{sh}}$	(d)	&$\epsilon$	&$P_{\rmn{prec}}$ (d)	&Fractional	&$q$\\
		&			&		&			&amplitude\\
\hline
1991	&0.107628(8)		&0.0568		&1.89			&0.20		&0.26\\
1993	&0.108567(2)		&0.0661		&1.64			&0.15		&0.31\\
1995	&0.10760(7)		&0.0566		&1.90			&0.13		&0.26\\
\hline
\end{tabular}
\end{table}

\begin{table}
\caption{Sidebands of the orbital period harmonics}
\label{harmonics}
\begin{tabular}{ccccc}
&\multicolumn{2}{c}{$\Omega=2\Omega_{orb}-\Omega_{prec}$}&\multicolumn{2}{c}{$\Omega=3\Omega_{orb}-\Omega_{prec}$}\\
&Pred. P (d) &Meas. P (d) & Pred. P (d) & Meas. P (d)\\
\hline
91	&0.052327(1)	&0.05241(1)	&0.0345661(6)	&0.034617(1)	\\
93	&0.0525477(4)	&0.052540(1)	&0.0346623(1)	&0.03466(1)	\\
95	&0.05232(1)	&0.052493(8)	&0.034563(6)	&0.034588(2)	\\
\hline
\end{tabular}
\end{table}

\subsection{The orbital parameters}
\label{qi}

The mass ratio, $q$, of a superhumper can be estimated, given its
period excess, $\epsilon$ (Patterson 1998b).
\[
\epsilon=\frac{0.23q}{1+0.27q}.
\]
This leads to the estimates of $q$ for V348 Pup shown in Table
\ref{shper}. We favour the mass ratio, $q = 0.31$, estimated from the
most accurate 1993 superhump period. We note that SPH simulations of
eccentric discs (Murray 1998, 1999) suggest a more complicated
relationship between $\epsilon$ and $q$ with the disc precession rate
depending on the gas pressure and viscosity of the disc in addition to
the mass ratio. This does not affect our substantive results.

If we assume that the secondary star is Roche lobe filling, the width,
$w$, of eclipse of a point source at the centre of the compact object
uniquely defines orbital inclination, $i$, as a function of $q$. We
can thus compute $i$ as a function of $q$ and $w$.

\begin{figure}
\psfig{file=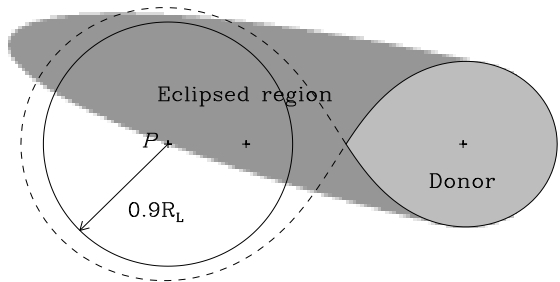,width=\columnwidth}
\caption{The eclipsed region of the orbital plane at phase $\phi_1$
showing how approximately half of the area of a $0.9R_L$ disc is
eclipsed at this phase.}
\label{qimethod}
\end{figure}

When the centre of the compact object (point \emph{P}) is first
eclipsed (orbital phase $\phi_1$), about half of the disc area will be
eclipsed, and therefore for a disc whose intensity distribution is
symmetric about the line of centres of the two stars, the fraction of
disc flux eclipsed at this phase will be $\sim 0.5$ (Figure
\ref{qimethod}). Similarly at the end of the eclipse of \emph{P}
(orbital phase $\phi_2$) the fraction of disc light visible is again
$\sim 0.5$. Further assuming that the lightcurve consists purely of
emission from a disc in the orbital plane, the full width of eclipse
at half intensity will be equal to $w$.

Using the average eclipses from each of our datasets to give $w$, we
obtained $i$. We checked the assumption that half of the disc area is
eclipsed at $\phi_1$ and $\phi_2$: for these values obtained it is a
good assumption. We therefore adopt orbital parameters $q = 0.31$, $i
= 81\fdg 1 \pm 1^{\circ}$ The conclusions drawn later are identical to
those obtained using $q = 0.36$ with corresponding $i = 80\fdg 0 \pm
1^{\circ}$ which resulted from an earlier estimate of $q$.

\subsection{The superhump modulation}
\label{superhumpmod}

\begin{figure*}
\psfig{file=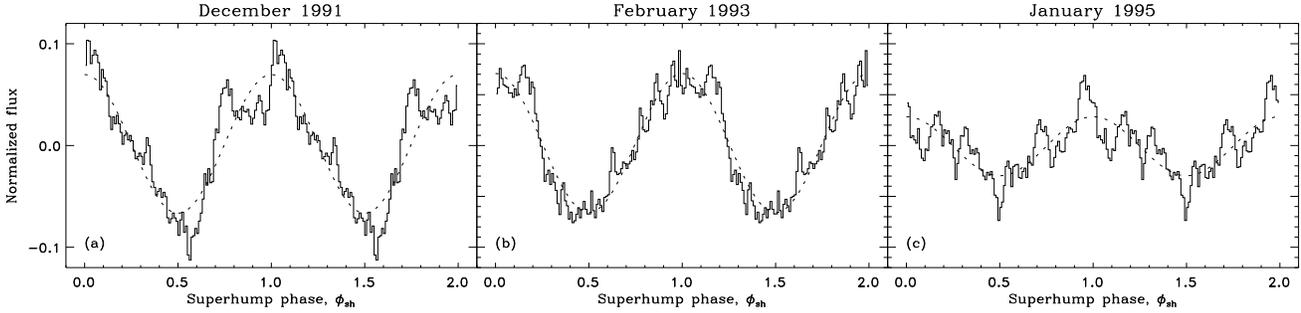,width=\textwidth}
\caption{The superhump modulations. For each dataset, the average 
orbital light curve was subtracted and the resulting data folded and
binned on superhump phase. Dotted lines show the sine waves fitted to
determine phasing}
\label{superhumps}
\end{figure*}

To define a zero point in superhump phase each set of data was folded
onto its detected superhump period, binned into 100 phase bins, and a
sine-wave was fitted to each (see Figure \ref{superhumps}).

We assessed the contribution of flickering to these curves by using
various methods of binning the data. Figure
\ref{superhumps} shows the curves produced from simply averaging the
points in each bin. Since flickering in the lightcurve consists of
brief increases in luminosity, by giving more weight to the lower
values in each bin or using only the lower points in a bin, the impact
of flickering on these superhump phase binned lightcurves should be
reduced. We therefore generated curves by averaging only the lowest 25
per cent of fluxes in each bin, and by weighting the lower values more
strongly than the higher values. The curves produced by these
different methods are virtually identical, except for a flux offset
between each. This suggests that flickering has little effect on the
superhump curves. Since we have extensive datasets and the timescales
of the flickering and the superhumps are very different, this is not
unexpected.

For a high inclination system a modulation on the superhump period
will arise due to the eclipses of the precessing accretion disc
changing as the disc orientation changes. This effect will occur in
addition to the intrinsic variations in luminosity which are observed
in non-eclipsing systems. Superhump modulations calculated using only
non-eclipsing phases are almost identical to those in Figure
\ref{superhumps}, which implies that in this system the form 
of the superhump light curve is not affected by the changing eclipse
shape.

The broad form of the modulation is consistent for all three sets of
observations. The peak-to-peak fractional amplitudes are shown in
Table \ref{shper} and decline steadily from year to year. The more
detailed structure, particularly the region with lower flux between
$\phi_{sh} \sim 0.8$ and 1.0 in the 1991 modulation, appears to be
genuine; we find neither flickering nor eclipses have significant
effect. Furthermore the abrupt changes do not correspond to changes in
system brightness from one night to the next.

Simpson \& Wood (1998) calculate pseudo-lightcurves for
superhumps. Assuming the light emitted from the disc is proportional
to changes in the total internal energy of the gas in the disc, they
present superhump shapes calculated for mass ratios of 0.050, 0.075
and 0.100 (their Figure 5). These curves have significant differences
in morphology: the $q=0.050$ curve has a sharp rise and slow decline,
the $q=0.075$ curve is reasonably symmetric, and the $q=0.100$ curve
has a slow rise and steeper decline.  The cleanest and most reliable
of our superhump curves, that from 1993, also shows an asymmetric
shape, with a slow rise and sharper decline, agreeing best with their
highest value of $q=0.100$. We expect $q \sim 0.31$ so this is
encouraging.

\subsection{Average lightcurves}
\label{averages}

\begin{figure}
\psfig{file=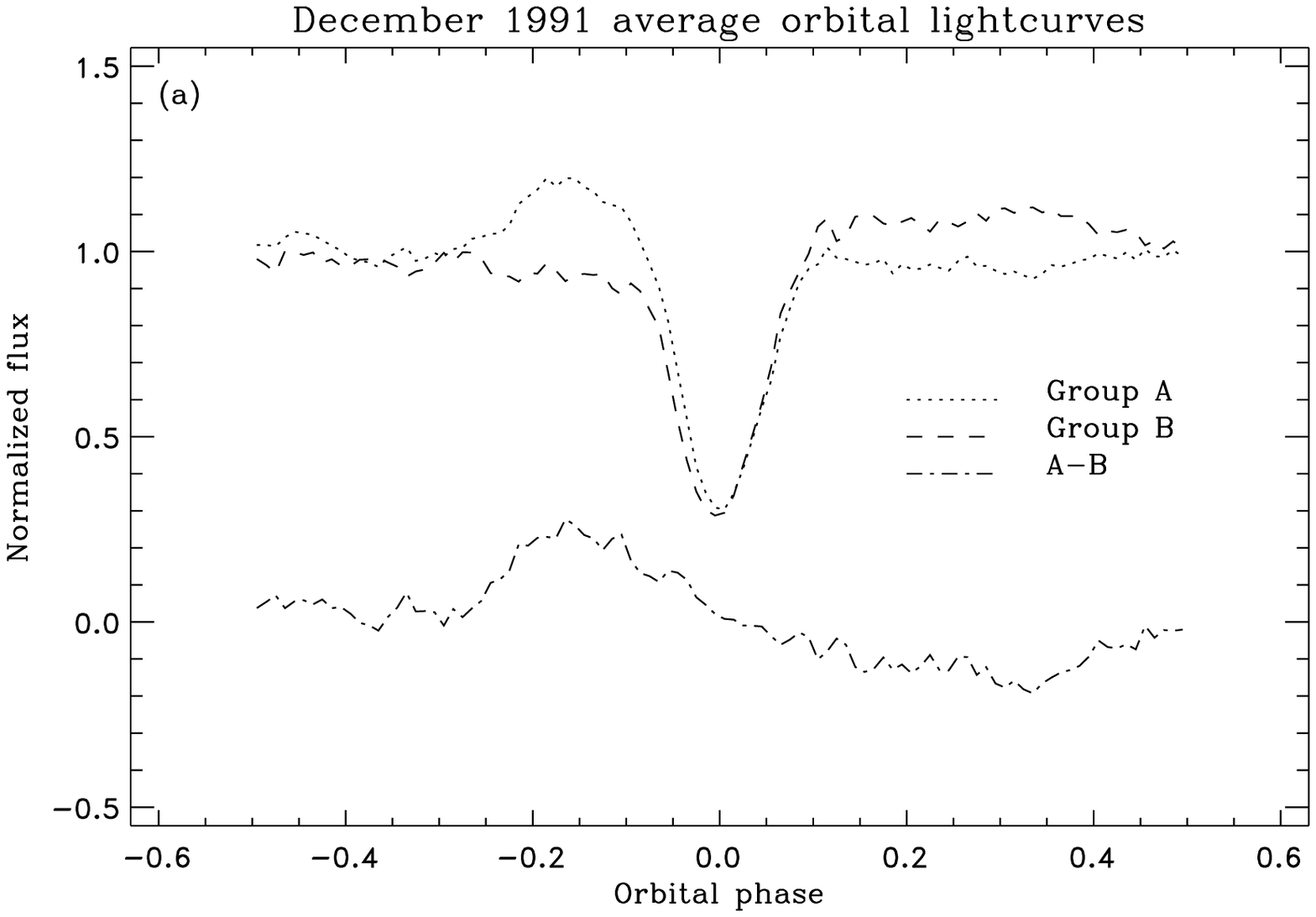,width=\columnwidth}
\psfig{file=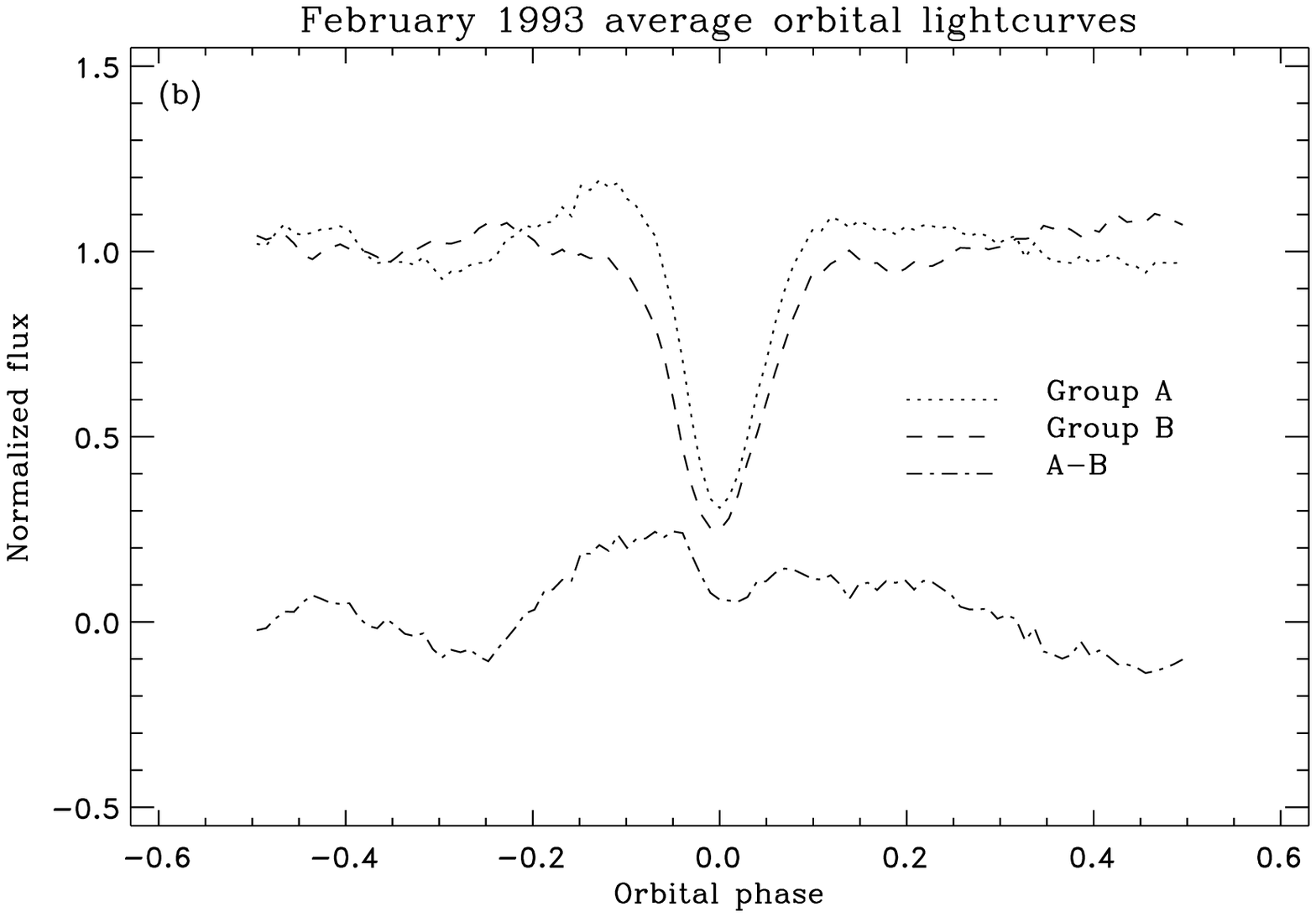,width=\columnwidth}
\psfig{file=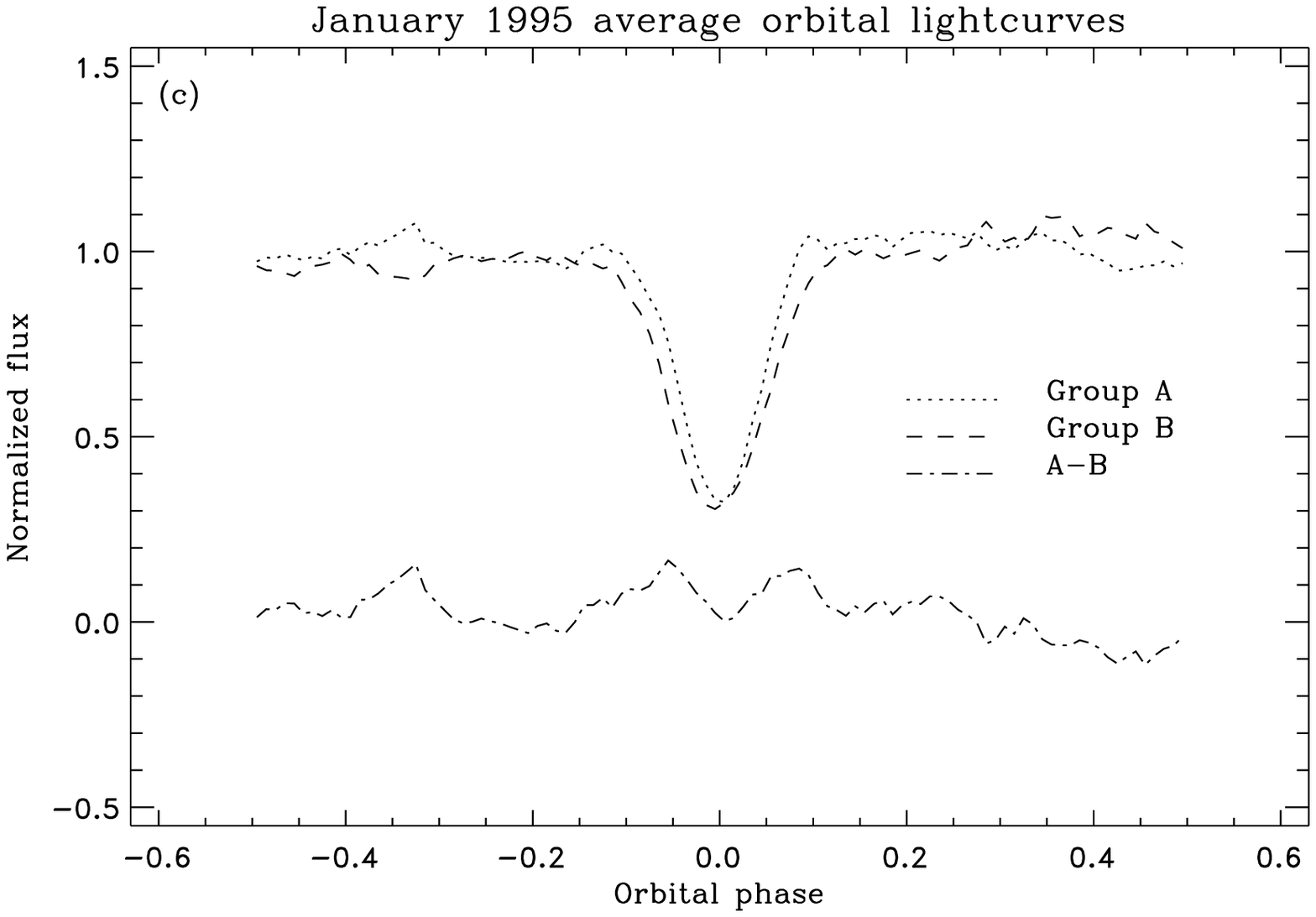,width=\columnwidth}
\caption{Average orbital lightcurves for each year. The data for
each year are sorted into two groups, so that all eclipses in each
group have mid-eclipse times with a similar superhump phase. The
values of $\phi_{\rmn{sh}}$ for each group are shown in Table
\ref{groups}.  The third curve in each plot is the difference between
the two lightcurves.}
\label{orbcurves}
\end{figure}

Because of the sampling of our data (Table \ref{obslog}), the observed
eclipses for each year appear in two rough groups, with superhump
phases separated by about 0.5. We compared the average orbital
lightcurves corresponding to each group. The average mid-eclipse
superhump phase, $\overline{\phi_{\rmn{sh}}}$, of each group, and the
range as indicated by the variance of ${\phi_{\rmn{sh}}}$ for each
group are shown in Table \ref{groups}, the average orbital curves are
shown in Figure
\ref{orbcurves}.

\begin{table}
\caption{Groups}
\label{groups}
\begin{tabular}{ccccccc}
&\multicolumn{2}{c}{December 1991}&\multicolumn{2}{c}{February
1993}&\multicolumn{2}{c}{January 1995}\\
&\multicolumn{2}{c}{$P_{\rmn{prec}}=1.89
d$}&\multicolumn{2}{c}{$P_{\rmn{prec}}=1.64
d$}&\multicolumn{2}{c}{$P_{\rmn{prec}}=1.90 d$}\\
Group&$\overline{\phi_{\rmn{sh}}}$&$\sigma_{\phi}$&$\overline{\phi_{\rmn{sh}}}$&$\sigma_{\phi}$&$\overline{\phi_{\rmn{sh}}}$&$\sigma_{\phi}$\\
\hline
A&0.24	&0.06	&0.02	&0.07	&0.03	&0.06	\\
B&0.71	&0.07	&0.48	&0.11	&0.54	&0.04	\\
\hline
\end{tabular}
\end{table}

The group A curve for 1991 (Figure \ref{orbcurves}a) displays a clear
hump at orbital phase around -0.2: the mid-eclipse superhump phase of
group A is 0.24, so eclipse should occur 0.25 in orbital phase after a
superhump maximum. Group B has mid-eclipse superhump phase 0.71,
meaning that the eclipse should occur around 0.3 in orbital phase
before the superhump. The post-eclipse flux in group B is higher than
in A, though the hump is not so sharp as in A.

Group A in the 1993 data displays a hump peaking just before
mid-eclipse, while the curve for group B is rather flat out of
eclipse. The mid-eclipse superhump phase of group A is 0.02, while
group B has superhump phase 0.48.  This is again consistent with the
position of the superhump. The difference curve (i.e. A-B) seems to
show a broad superhump which is partially eclipsed by the secondary.
We expect group B to display a hump around orbital phase 0.5. This is
not obvious, meaning that the superhump is more prominent when is
occurs at orbital phase 0 (group A) than at phase 0.5. The superhump
light is therefore not emitted isotropically\footnote{Eclipses of the
superhump light will obviously be most important when the secondary is
at inferior conjunction, i.e. when the superhump maximum is at orbital
phase 0}.  Schoembs (1986) also noted a similar effect in OY Car. This
will be considered further in Section
\ref{discussion}.

The superhump phasing of the 1995 eclipses is almost the same as those
observed in 1993, but the smaller extent of the 1995 data and the low
fractional amplitude makes identifying the hump difficult without
first subtracting the orbital light curve (compare Figs
\ref{superhumps}b and \ref{superhumps}c). However, the flux during
eclipse for group A is higher than for group B, consistent with the
phasing which suggests that a superhump should occur at mid-eclipse.

\subsection{Eclipse parameters}
\label{ecpar}

\begin{figure*}
\psfig{file=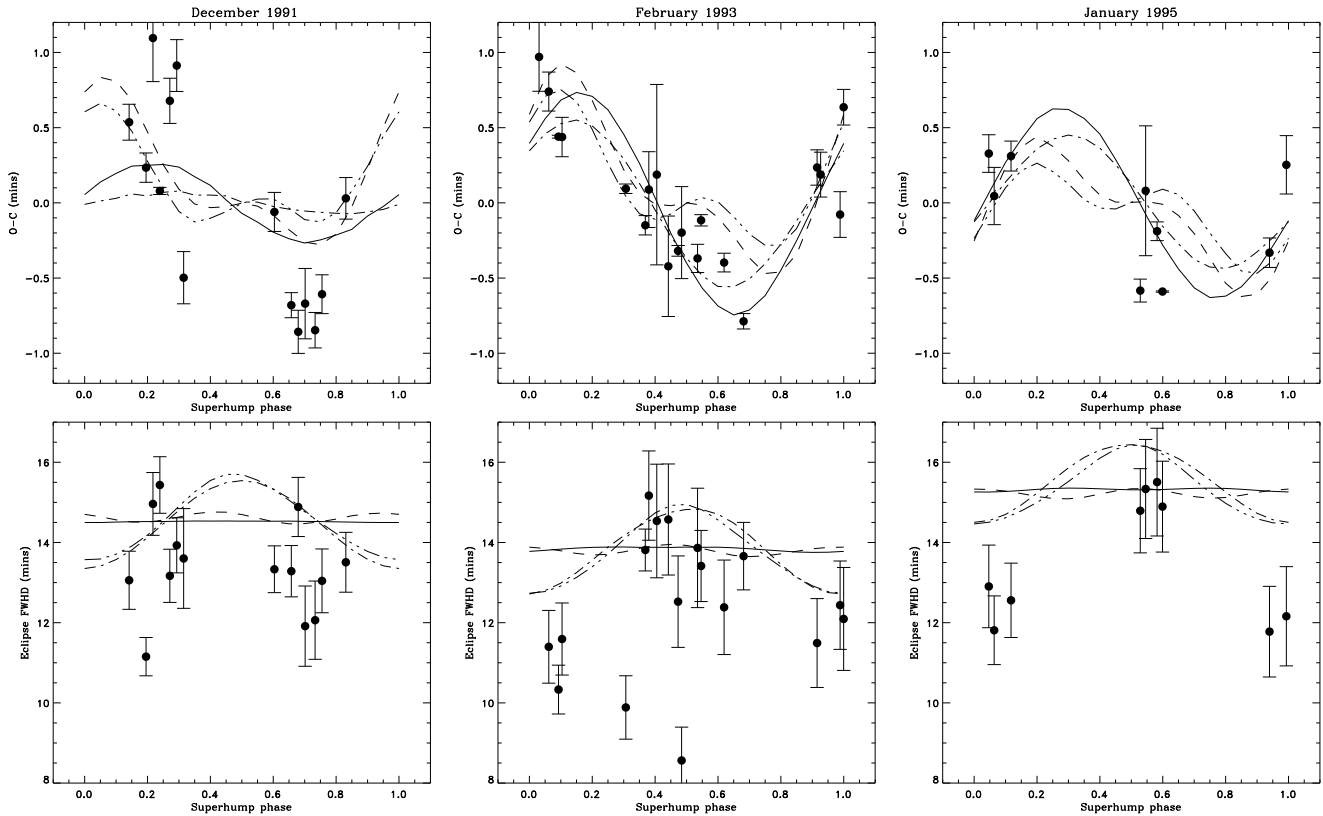,width=\textwidth,angle=90}
\caption{Eclipse parameters. The O-C mid-eclipse times and the full 
eclipse widths at half depth. The superhump phase, $\phi_{\rmn{sh}}$,
is calculated using the superhump period, $P_{\rmn{sh}}$ determined
directly from the corresponding dataset. The continuous, dashed,
dot-dashed and dot-dot-dot-dashed curves correspond to predictions
from the lightcurve fits {\emph f}, {\emph e}, {\emph a} and {\emph b}
described in Section
\ref{ecpar}.}
\label{eclipsestuff}
\end{figure*}

The O-C mid-eclipse times and the eclipse widths are shown in Figure
\ref{eclipsestuff}. The mid-eclipse times were determined both by
fitting a parabola to the deepest half of each eclipse and also by
finding the centroid. The discrepancy between the two determinations
provides an indication of the uncertainty.

As the eccentric disc precesses slowly in our frame, we expect to see
the eclipse width and midpoint phase modulated on the apsidal
precession period. These quantities will be
similarly modulated in superhump phase, since the superhump phase and
precession phase of an eclipse are both measures of the relative
orientation of disc and secondary star at mid-eclipse. Figure
\ref{eclipsestuff} shows that eclipse timings for all years
exhibit a precession period modulation.  The widths also show evidence
of a modulation. The limited superhump phase coverage means that
conclusions cannot easily be drawn from inspecting the datapoints
alone. Such variations in eclipse asymmetry have been observed in
other superhumping systems e.g. OY Car (Schoembs 1986) and Z Cha
(Warner and O'Donoghue 1988, Kuulkers et al. 1991).  To further
investigate the disc shape we produced a simple model which was then
fitted to the observed lightcurves.

Our simple eccentric disc prescription has a circular inner boundary
with radius $r_{min}$, centred on the white dwarf. The outer boundary
is an ellipse of semi-major axis $a_{max}$, eccentricity $e$, with one
focus also centred on the white dwarf. The disc brightness at distance
$r(\alpha)$ from the white dwarf at an angle $\alpha$ to the
semi-major axis is

\[
S(\alpha)\propto\left(\frac{r(\alpha)-r_{min}}{r_{max}(\alpha)-r_{min}}+\frac{r_{min}}{a_{max}(1-e)-r_{min}}\right)^{-n}
,\] where $r_{max}(\alpha)$ is the distance from the white dwarf to
the outer disc boundary at orientation $\alpha$. Brightness contours
are therefore circular at the inner boundary, smoothly changing to
elliptical at the outer boundary. This form for $S(\alpha)$ reduces to
$S\propto r^{-n}$ if the disc is circular. Our model is sensible for a
tidally distorted disc, since the tidal influence of the secondary
star is unimportant at the inner disc, so we expect a more or less
circular inner disc.

In an inertial frame, the disc slowly precesses progradely with period
$P_{prec}$. With respect to the corotating frame of the system, this
disc then rotates retrogradely with period $P_{\rmn{sh}}$. Let the
relative orientation of the line of apsides of the disc with the line
of centres of the binary when superhump maximum occurs be
$\phi_{disc}$. The structure of the disc in our model is therefore
described by five parameters : $r_{min}$, $a_{max}$, $e$, $n$ and
$\phi_{disc}$.  This model was chosen as the simplest way to model an
eccentric precessing disc, with as few parameters as possible.  A
similar model was used by Patterson, Halpern \& Shambrook (1993) to
model the disc in AM CVn.

We generated synthetic lightcurves for eclipses of our model disc
using the orbital parameters from Section \ref{qi}. By varying the
parameters of our model to minimize the reduced $\chi_r^2$ of the fit,
we obtained a best fit of our model to our lightcurves. The smoothed
superhump was subtracted from the lightcurves before fitting the model
in order to remove the intrinsic variation in the disc flux, enabling
us to study the shape of the disc. In Section \ref{averages} we noted
that the superhump is more visible for $\phi_{sh}=0$ at
$\phi_{orb}=0$, so ideally we should subtract a superhump modulation
which takes account of this variation in superhump prominence, but
insufficient sampling of the disc precession phase by our data
prevents us from doing this.

The downhill simplex method for minimizing multidimensional functions
was used (the {\sevensize AMOEBA} routine from Press et al.,
1996). Each orbit (centred on an eclipse) was allowed to have a
different total disc flux and a different uneclipsed flux. This
prevents the variation of $\sim$10 per cent in flux from one orbit to
the next from interfering with the results, and allows for the
possibility of a contribution to the lightcurve from regions never
eclipsed by the secondary star. The errors in the fluxes were
estimated as being $\sigma\sqrt{flux}$ where $\sigma$ is the square
root of the variance of the flux between orbital phase 0.2 and
0.8. This estimate therefore includes the effect of flickering.

We used two methods to assess the robustness and accuracy of these
fits. Monte-Carlo methods estimated the size of the region in
parameter space which has a $\sim$75 per cent chance of containing the
`true' values of the parameters. To assess the uniqueness and
robustness of each solution we carried out the fitting process 20
times for each model/data combination starting each fit with a
different random initial simplex. The solution chosen is that with the
lowest value of reduced $\chi^2$. Extreme outlier solutions are
rejected and the variance in the parameters for the remaining
solutions is a measure of the accuracy with which the {\sevensize
AMOEBA} routine converges to a unique solution. The errors quoted in
Table \ref{fits} for each parameter are whichever is the greater of
confidence region estimate or the variance in the parameter from the
multiple fits. The parameters resulting from all the fits are shown in
Table \ref{fits}. $R_L$ is the Eggleton radius of the primary Roche
lobe (Eggleton 1983). The $\chi^2$ surface in parameter space is not
perfectly smooth. There is a broad global minimum superimposed with
smaller amplitude bumps. Close to the global minimum the gradient of
$\chi^2$ is low and so small bumps can lead {\sevensize AMOEBA} to
settle into a local minimum near the real minimum.

Fits using the model as described above will be referred to as fit
{\emph f}. The emission extends out to 80 -- 90 per cent of Roche lobe
size, while the largest radius of the disc is $a_{max}(1+e) = 0.97
R_L$ for 1995 data. These results suggest that the disc does indeed
extend out to the tidal cut-off radius, $r_{tide}$; $r_{tide} \sim
0.9R_L$ (Paczynski 1977).

The 1995 radius is 10 per cent larger than the 1991 and 1993
radii. The 1991 and 1993 observations are in blue CuSO$_4$ filter
light, which should come from hotter inner disc regions; the 1995
observations are in R which is expected to weight the outer disc
emission more heavily, perhaps causing the inferred disc radius to be
largest for the 1995 observations. A simple model black-body disc with
a $T \propto R^{-\frac{3}{4}}$ temperature distribution, with
$T=10,000K$ at the inner disc radius $r_1=0.1R_L$ and with an outer
disc radius $r_2=R_L$ was used to calculate eclipse profiles in the R
and B band. Fitting model {\emph f} to these eclipses showed no
significant difference in outer disc radius between the R band and
CuSO$_4$ filter fits.

The inner disc boundary and the index $n$ in the flux distribution are
poorly constrained. The eccentricity is robustly non-zero; the
changing eclipse shape demands a non-axisymmetric disc.

The most interesting result is that all three datasets have
$\phi_{disc}$ around 0.4 to 0.5. This means that in our elliptical
model the secondary star sweeps past the \emph{smallest} radius part
of the disc at superhump maximum - a result unexpected if tidal
stressing of the disc by the gravitational influence of the secondary
star is responsible for the superhump light. The implications of this
result are discussed later.

We adjusted the model so that the eccentricity varied during the
superhump cycle as $e(\phi_{sh})=e_0cos^2\pi\phi_{sh}$. This will be
referred to as fit {\emph e}. This variation in eccentricity follows
Simpson \& Wood's (1998) simulation in which the disc varies between
being highly eccentric at the superhump maximum to almost circular
away from the superhump. The results for $r_{min}$ and $a_{max}$
change very little, with $a_{max}$ again larger in the red (1995) than
the blue (1991 and 1993). $\phi_{disc}$ is unchanged from fit {\emph
f} within the errors for all three years. The maximum eccentricity,
$e_0$, is larger than when $e$ was constant. This is expected since
the eccentricity is demanded by the variation in O-C mid-eclipse
times, and these O-C times are non-zero at times when $e$ is less than
$e_0$.

Next, we obtained fits in which the eccentricity was again constant,
but where $a_{max}$ was allowed to vary between $a_1$ at superhump
maximum to $a_2$ half a superhump period later;
$a_{max}(\phi_{sh})=a_1+(a_2-a_1)sin^2\pi\phi_{sh}$. This will be
referred to as fit {\emph a}. This was an attempt to reproduce the
observed variations in eclipse width (Figure
\ref{eclipsestuff}). $r_{min}$, $e$ and $\phi_{disc}$ are essentially
the same as in the first fit, while the values of $a_1$ and $a_2$
imply a variation in $a_{max}$ of amplitude 20 -- 25 per cent in the
blue and 14 per cent in the red, with the disc being smallest at
superhump maximum. At its largest, the disc extends to the edge of the
Roche lobe. While these implied variations in disc size are large,
they are comparable to the uncertainties in $a_1$ and $a_2$, and so
must be treated with caution.

The final variation on our model was to allow both $e$ and $a_{max}$
to vary as described above. This will be referred to as fit {\emph
b}. The eccentricity and $\phi_{disc}$ are little different from the
fits with $e$ varying periodically and $a_{max}$ constant, while the
values of $a_1$ and $a_2$ follow those from the previous fit ($e$
constant and $a_{max}$ varying).

\begin{table}
\begin{center}
\caption{Treatment of $a_{max}$ and $e$ in our four models. $r_{min}$, $n$ and $\phi_{disc}$ are constant in all four models.}
\label{models}
\begin{tabular}{ccccc}
Fit &$a_{max}$ &$e$\\
\hline
{\emph f}	&	constant	&	constant\\
{\emph e}	&	constant	&	$e_0cos^2\pi\phi_{sh}$\\
{\emph a}	&	$a_1+(a_2-a_1)sin^2\pi\phi_{sh}$	&	constant\\
{\emph b}	&	$a_1+(a_2-a_1)sin^2\pi\phi_{sh}$	&	$e_0cos^2\pi\phi_{sh}$\\
\hline
\end{tabular}
\end{center}
\end{table}

\begin{figure}
\psfig{file=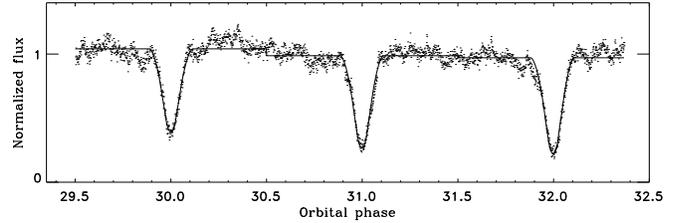,width=\columnwidth}
\caption{One night of the January 1995 data with the best fitting lightcurve using model {\emph a} plotted as a continuous line. The discontinuities in the fit illustrate the different disc fluxes and uneclipsed fluxes allowed for each orbit. See Section \ref{ecpar}.}
\label{examplefit}
\end{figure}

The treatment of $a_{max}$ and $e$ for each fit is summarized in Table
\ref{models}.  In Figure \ref{examplefit} we show part of the fit to
the 1995 dataset using model {\emph a}. The fit (shown as the
continuous line) shows the different disc flux and uneclipsed flux
allowed by our model in the form of the discontinuities at phase 30.5
and 31.5. The level of flickering out of eclipse can also clearly be
seen, and was taken into account in our estimate of the errors as
described above.

\begin{figure}
\psfig{file=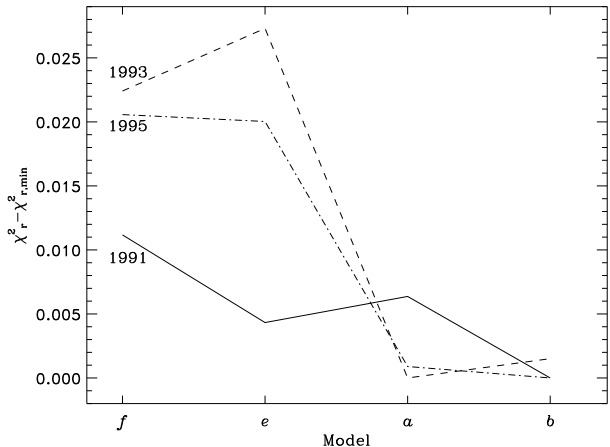,width=\columnwidth}
\caption{The minimum values of $\chi^2$ achieved for the various models
fitted (Section \ref{ecpar}).}
\label{comparefits}
\end{figure}

Formally, the best model is that which achieves the lowest value of
reduced $\chi^2$ ($\chi^2_r$). In Figure \ref{comparefits} we show the
values of $\chi^2_r$ achieved for each model and dataset relative to
the lowest. The minimum $\chi^2_r$ achieved was around 0.8 for all
datasets and models. This figure shows that the fits {\emph a} and
{\emph b}, i.e. those in which $a_{max}$ varies on the superhump cycle,
produce significantly better fits to the 1993 and 1995
observations. The variation of $e$ during the superhump cycle has
little effect on the quality of these fits. There is less significant
difference between the $\chi^2$ achieved by the different fits to the
1991 observations, although allowing $a_{max}$ or $e$ to vary during
the superhump cycle produces a better fit than when they are both
constant. The significant reduction in $\chi^2_r$ achieved by allowing
$a_{max}$ to vary implies that this model best represents the
behaviour of the system.

It is also interesting to compare how well each model predicts the
variation in the eclipse width and O-C mid-eclipse times. The
predictions of each model are plotted in Figure \ref{eclipsestuff}. It
is periodic variation in these two eclipse characteristics which
requires the disc to be eccentric. The models poorly reproduce the O-C
variations in the 1991 data. While the phasing of the predicted
variation agrees with the observations, the amplitude is too low. The
fits in which $e$ varies during the superhump cycle predict a larger
modulation in O-C times, a result of the larger eccentricity in these
fits, but the agreement for these fits is still poor. The 1991
lightcurves suffer more from flickering than the 1993 and 1995 data,
with many eclipses distorted as a result. This is the most likely
explanation for the poor agreement between our model and the 1991
lightcurves. The agreement between the predicted and observed O-Cs is
very good for all models for the 1993 observations. The variation in
eclipse width is only reasonably modelled by those fits in which
$a_{max}$ varies. The same is true of the 1995 fits.

The result of these comparisons between the different models, both the
formal comparison of reduced $\chi^2$ and the more subjective
`chi-by-eye' considerations of the O-C times and eclipse widths is
that the models in which $a_{max}$ varies during the superhump cycle
predict the observations better than those in which $a_{max}$ is
constant.

All four models agree on three important points. The values of
$a_{max}$, $a_1$ and $a_2$ show that the disc is large, filling at
least about 50 per cent of the Roche lobe area. The disc is not
axisymmetric. From the consistent values of $\phi_{disc}$ we see that
when the superhump reaches maximum light, the light centre of the disc
is on the far side of the white dwarf from the donor star.

\begin{table*}
\begin{center}
\begin{minipage}{110mm}
\caption{Parameters resulting from fitting various models to the lightcurves (see Section \ref{ecpar}). For the behaviour of $a_{max}$ and $e$ in each model see Table \ref{models}.}
\label{fits}
\begin{tabular}{cccccccc}
Fit & Year & $r_{min}/R_L$ &\multicolumn{2}{c}{$a_{max}/R_L$}& $e$ & $n$ & $\phi_{disc}$\\ & & & $a_1/R_L$ & $a_2/R_L$ & $e_0$ & &\\
\hline
{\emph f} & 1991 &0.13(13) & \multicolumn{2}{c}{0.81(10)} &0.035(03) &0.61(81) &0.46(17)\\
{\emph f} & 1993 &0.12(09) & \multicolumn{2}{c}{0.83(07)} &0.106(02) &0.94(65) &0.41(11)\\
{\emph f} & 1995 &0.11(13) & \multicolumn{2}{c}{0.90(09)} &0.082(03) &0.64(71) &0.53(13)\\
\\
{\emph e} & 1991 &0.13(10) & \multicolumn{2}{c}{0.82(08)} &0.121(06) &0.62(76) &0.35(15)\\
{\emph e} & 1993 &0.09(10) & \multicolumn{2}{c}{0.81(07)} &0.144(03) &0.67(60) &0.41(03)\\
{\emph e} & 1995 &0.08(10) & \multicolumn{2}{c}{0.88(06)} &0.102(04) &0.48(52) &0.55(15)\\
\\
{\emph a} & 1991 &0.12(14) &0.72(12) &0.88(17) &0.029(02) &0.59(84) &0.51(22)\\
{\emph a} & 1993 &0.11(11) &0.76(08) &0.99(16) &0.099(04) &1.11(73) &0.41(15)\\
{\emph a} & 1995 &0.10(16) &0.78(11) &0.91(09) &0.068(05) &0.34(96) &0.53(21)\\
\\
{\emph b} & 1991 &0.12(15) &0.74(09) &0.89(20) &0.104(08) &0.55(79) &0.35(18)\\
{\emph b} & 1993 &0.10(09) &0.76(08) &0.98(13) &0.140(05) &0.99(61) &0.41(12)\\
{\emph b} & 1995 &0.09(17) &0.80(09) &0.93(10) &0.084(06) &0.42(61) &0.56(22)\\
\hline
\end{tabular}
\end{minipage}
\end{center}
\end{table*}

\subsection{Eclipse mapping}
\label{eclipsemapping}

In Section \ref{ecpar} we used the changing eclipse profiles to
constrain the parameters of a model intensity distribution. An
alternative method for investigating the distribution of emission in
the orbital plane is the commonly used eclipse mapping technique
developed by Horne (1985). This method assumes that intensity
distribution is fixed in the corotating frame of the binary, lies flat
in the orbital plane and is constant. The surface of the secondary
star is described by its Roche potential surface. Maximum entropy
methods are used to obtain the intensity distribution by comparing the
calculated lightcurve and observed lightcurves. The $\chi^2$ statistic
is used to ensure consistency between the observations and the fitted
distribution, while the entropy is used to select the most appropriate
solution from the multitude of possibilities. The entropy is usually
defined such that the final solution is the smoothest or most
axisymmetric map consistent with the observations. This technique has
been widely used, and O'Donoghue (1990) employed it to locate the
source of the strong normal superhumps in Z Cha.

In order to study the shape of the precessing disc in V348 Pup, the
{\sevensize PRIDA} eclipse mapping code of Baptista \& Steiner (1991)
was modified so that the intensity distribution was fixed in the
precessing disc frame rather than the corotating frame of the
binary. Each year's data was split into two groups (Section
\ref{averages}) but the lightcurves were not folded on orbital
phase. This enabled us to obtain two maps for each year, corresponding
to the groups given in Table \ref{groups}. Since we expect the
intensity distribution to change throughout the superhump cycle,
grouping the eclipses as described means that the intensity
distribution should be roughly the same for all eclipses in a group,
an assumption of the eclipse mapping method. The superhump modulation
was subtracted from each lightcurve as in Section
\ref{ecpar}. Normalization of the lightcurves was achieved by using
the values for total disc flux and uneclipsed flux for each orbit
obtained during the fitting procedure in Section \ref{ecpar}. The
uneclipsed flux was subtracted from each orbit and fluxes were then
rescaled to produce an effective disc-only lightcurve. Various other
normalization techniques were tested, and the detail of the
reconstructed maps was sensitive to these changes. We used orbital
parameters $q=0.31$ and $i=81^\circ$, and looked for the most
axisymmetric solution consistent with the data.

\begin{figure}
\psfig{file=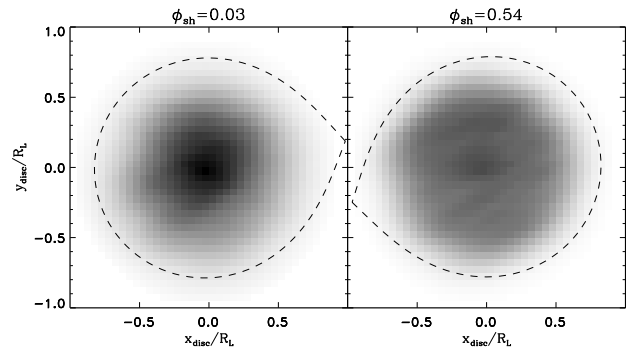,width=\columnwidth}
\caption{Maximum entropy eclipse maps of the intensity distribution in the \emph{precessing} frame of the disc from the 1995 data. The \emph{average} orientation of the primary Roche lobe for each group is also shown. See Section \ref{eclipsemapping}. The grey scale is the same for both maps.}
\label{eclipsemaps}
\end{figure}

\begin{figure*}
\psfig{file=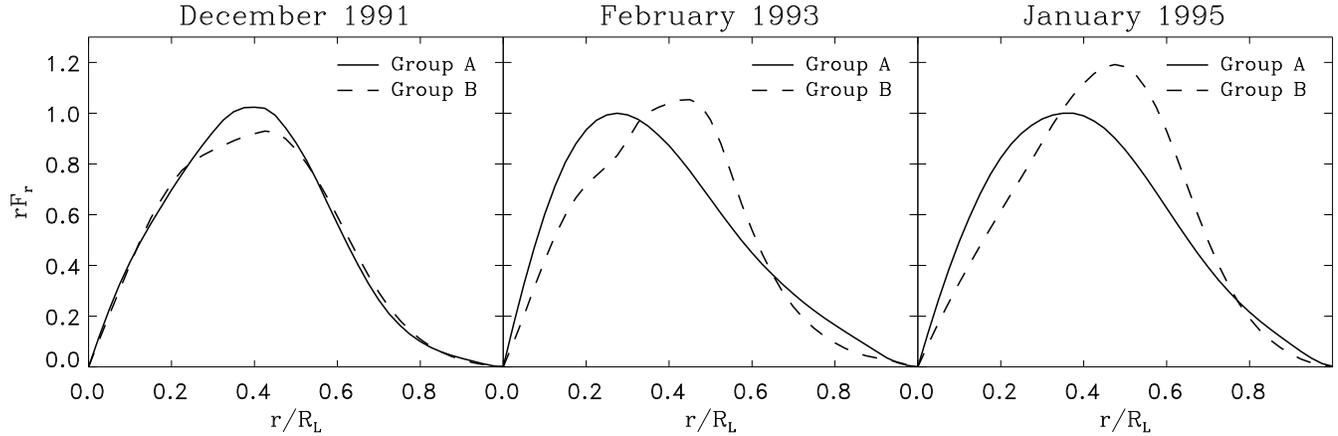}
\caption{Azimuthally averaged flux in maximum entropy eclipse maps. Quantity plotted, $rF_r$, is proportional to flux in annulus at radius $r$. See Section \ref{eclipsemapping}.}
\label{mapprofiles}
\end{figure*}

The most consistent result revealed by these eclipse maps is that the
emission at superhump phase 0.5 is less centrally concentrated than at
superhump phase 0. This is illustrated in Figure \ref{eclipsemaps}
which shows the maps for the two 1995 groups of eclipses. Figures
\ref{mapprofiles}a to \ref{mapprofiles}c show the azimuthally averaged
brightness distribution of each map. They show the flux at radius $r$,
$F_r$, multiplied by $r$; this quantity is proportional to the total
flux in an annulus at radius $r$. Figures \ref{mapprofiles}b and
\ref{mapprofiles}c show how the disc extends further out at superhump
phase 0.5 than at phase 0, while the curves in Figure
\ref{mapprofiles}a are both nearly the same, as expected since both of
these curves show the situation roughly half way between superhump
phase 0 and 0.5. This result is in agreement with the results of our
fits of model {\emph a} in which the disc size was allowed to
change. These fits showed that the size of the emission region is
larger at superhump phase 0.5 than at phase 0. The maps are
asymmetric, but due to the sensitivity described above, we draw no
conclusions from the detailed structure.

\section{Discussion}
\label{discussion}

The phase of the superhump relative to the conjunction of the line of
centres of the system and the semi-major axis of the disc should make
it possible to determine whether the bright spot model or the tidal
heating model better explains the source of the superhump. The
simplest tidal model predicts that the superhump light should peak
when (or slightly after) the largest radius part of the disc coincides
with the line of centres. This is because the tidal interaction is
strongly dependent on distance from the secondary, and so will be most
significant in regions where the disc extends out close to the L1
point.  However, if the bright spot model is to be believed, then the
the superhump light source will be brightest when the accreting
material has the furthest to fall. In other words, the superhump
should occur when the stream impacts on the disc at its smallest
radius.

\begin{figure}
\psfig{file=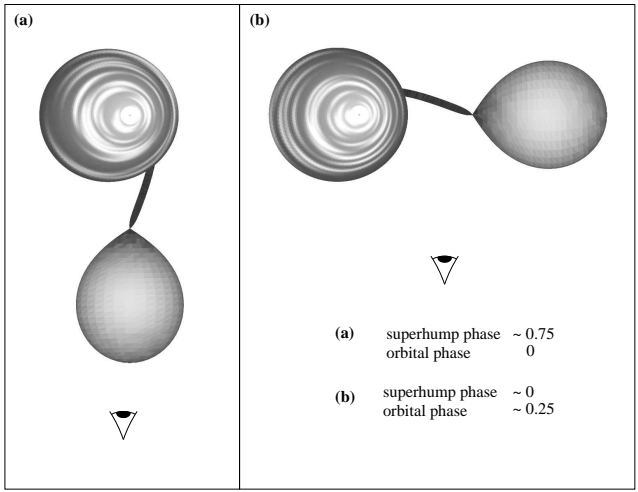,width=8.4cm,angle=270}
\caption{The first row of Figure \ref{eclipsestuff} shows
the eclipses to be earliest at superhump phase $\phi_{\rm{sh}}\sim$
0.75, which implies orientation (a), from which we deduce the relative
phasing of disc and secondary star at superhump maximum
($\phi_{\rm{sh}}=0$) shown in (b).}
\label{phasing}
\end{figure}

The mid-eclipse times shown in the top row of panels in Figure
\ref{eclipsestuff} show the eclipses to be earliest around superhump
phase 0.75 in all cases. Assuming that the centre of light of the
eccentric disc is offset from the white dwarf in the direction of the
largest radius, we can deduce the disc orientation during these
eclipses to be as shown in Figure \ref{phasing}a. A quarter of a
superhump period later, the orientation of the disc has barely
changed, the secondary will be lined up with the smallest radius part
of the disc and the superhump phase will be 0.0 (Figure
\ref{phasing}b). Therefore superhump maximum occurs when the
secondary star is lined up with the smallest radius part of the disc.
The values of $\phi_{disc}$ in Table \ref{fits} agree with this
deduction. This phasing is consistent with the bright spot model for
the superhump emission but is inconsistent with the simple tidal
heating model. In Section \ref{averages}, we noted that the superhump
light appears not to be emitted isotropically: the superhump is
strongest when it occurs around orbital phase 0. This is easily
explained if the major contribution to superhump light is the bright
spot: the bright spot is most visible when it is on the nearside of
the accretion disc. Schoembs (1986) observed late superhumps in the
eclipsing SU UMa dwarf nova OY Car, also a high inclination
system. When a superhump was coincident with a pre-eclipse orbital
hump, the combined amplitude was greater than that predicted for a
linear superposition of the individual amplitudes i.e. OY Car's late
superhumps were strongest around orbital phase 0. However, van der
Woerd et al. (1988) studied the dwarf nova VW Hyi, concluding that
there was no correlation between the orbital phase and amplitude of
late superhumps. Since VW Hyi has an intermediate inclination,
$\sim60^\circ$ (Schoembs \& Vogt 1981), the bright spot visibility need
not vary with phase, so if the bright spot is the main superhump light
source we expect no variation in superhump amplitude with orbital
phase.

Krzeminski \& Vogt (1985) studied OY Car during a super-outburst and
through variations in the O-C eclipse timings deduced the presence of
an eccentric disc with phasing similar to that in V348 Pup. Krzeminski
\& Vogt's definition of O-C time was criticized by Naylor et
al. (1987), with Naylor et al. concluding that the O-C evidence was
weaker than originally thought.

Schoembs (1986) followed OY Car from early in a super-outburst almost
until the return to quiescence, observing the $\sim 180^{\circ}$ phase
change from normal superhumps around the height of the outburst to
late superhumps during the decline of the super-outburst. Patterson et
al. (1995) observed the same change in superhump phase late in a
super-outburst of V1159 Ori. Hessman et al. (1992) studied OY Car at
the end of a super-outburst. By looking at the varying hot spot
eclipse ingress times, and considering the trajectory of the accretion
stream, they concluded that the disc was eccentric. The orientation of
the disc at superhump maximum was very similar to that which we find
in V348 Pup.

The broad waveform of these late superhumps in OY Car (Hessman et al.)
was also similar to the superhump modulation in V348 Pup. Such
similarity between late superhumps in OY Car and the superhumps in
V348 Pup is not surprising. Late superhumps in dwarf novae appear
towards the end of the superoutburst, after the disc has had time to
adjust to its high state. V348 Pup is persistently in a high
state. Superhumps in a novalike system might resemble those to which
the superhumps in a superoutbursting dwarf nova would tend if it
remained in superoutburst for a long time. It seems likely that the
mechanism responsible for late superhumps in SU UMa systems is the
same mechanism responsible for superhumps in V348 Pup. However,
Skillman et al. (1998) observed strong superhumps in the nova-like TT
Ari throughout 1997 whose waveform is triangular like those of normal
superhumps in dwarf novae.

There are many other studies of the disc structure in SU UMa stars
during superoutburst. Vogt (1981) and Honey et al. (1988) found
evidence for an eccentric precessing disc in Z Cha from the radial
velocity variations of various absorption and emission lines
respectively. The very prominent normal superhump in Z Cha made it
possible for Warner \& O'Donoghue (1988) to study the location of the
superhump light source. They found strong departures from axisymmetry
in the superhump surface-brightness.  O'Donoghue (1990) employed a
modified eclipse mapping technique to Z Cha lightcurves and found the
normal superhump light coming from three bright regions of the disc
rim, located near the L1 point and the leading and trailing edges of
the disc, concluding that the superhumps are tidal in origin, and that
a highly eccentric disc with a smooth brightness distribution is not
necessary to explain superhump behaviour. One anomalous eclipse did
confine the superhump light source in Z Cha to the region of the
quiescent bright spot. van der Woerd et al. (1988) concluded that the
late superhumps in VW Hyi come from an optically thin plasma and could
be a result of tidal interaction.

In the SPH simulations of Murray (1996, 1998) pseudo-lightcurves are
produced by assuming the heat produced by viscous dissipation to be
radiated away where it is generated. Murray (1996, 1998) reveals an
extended superhump light source in the outer disc, while Murray (1996)
reveals an additional superhump modulation which arises from the
impact of the accretion stream with the edge of the disc occuring at a
varying depth in the primary Roche potential. This additional weaker
superhump modulation is approximately 180$^\circ$ out of phase with
the modulation due to tidal stressing, another similarity between late
superhumps in dwarf novae, the persistent superhumps in V348 Pup and
the bright spot model.

If we consider the stream to impact the disc at radius $r$ in a
$\frac{1}{r}$ potential, then the luminosity, $L$, of the bright spot
should vary roughly as $\Delta(\frac{1}{r})$. Considering the change
in $r$ as the disc with eccentricity $e$ precesses, we get
$\frac{\Delta L}{L} \sim 2e$. The eccentricities we find are in the
range 0.035--0.15 predicting superhump fractional amplitude in the
range 0.07--0.3. This is consistent with the measured superhump
amplitudes in V348 Pup (Table
\ref{shper}).

While we limit the conclusions drawn from our eclipse maps in Section
\ref{eclipsemapping}, there are a number of points deserving
consideration.

Our eclipse maps do not show evidence of a bright spot, but this does
not rule out the possibility that a bright spot is the source of the
superhump light, for the following reasons. First, we subtracted the
superhump modulation from the lightcurves before performing the
eclipse mapping, which should reduce the contribution of the bright
spot in the maps if it is the primary source of the superhump
light. Also, our maps are fixed in the precessing disc frame, rather
than the orbital frame of the system, so the hot spot should be
blurred azimuthally in our maps by $\sim70^\circ$ corresponding to the
eclipse width of the system of $\sim0.2$ in orbital phase. There will
be additional azimuthal blurring since the eclipses contributing to
each map have a spread of disc orientations at mid-eclipse
corresponding to the values of $\sigma_\phi$ in Table
\ref{groups}. Azimuthal structure in the maps is also suppressed by
looking for the maximally axisymmetric solution.

The eclipse maps tell us that the azimuthally averaged radial extent
of the emission is lowest at superhump maximum, shown in Figure
\ref{mapprofiles}. If this change in extent of the emission
region is interpreted as a result of a changing disc size, the the
smaller disc radius at superhump maximum is consistent with the bright
spot model for the superhump light source.

In the SPH models of Simpson \& Wood (1998), the symmetry axis of the
disc is aligned roughly perpendicular to the line of centres of the
system when the superhump reaches maximum intensity.  Inspecting their
plots suggests that this model would lead to eclipses being earliest
and widest at superhump phase 0 contrary to our findings. Simpson \&
Wood stress that their pseudo-lightcurves should be treated cautiously
since no radiative processes were explicitly considered.  The
difference in the mass ratio between V348 Pup and the values
considered in Simpson \& Wood's simulations may affect the phasing of
the early eclipses and superhump maximum, given that the predicted
superhump waveform is sensitive to $q$.  Furthermore, the spiral
density waves in their simulations complicate the structure, so that
simulated maps of the intensity may in fact produce reasonable
agreement with our findings.

SPH simulations (Murray 1996 \& 1998, Simpson \& Wood 1998) show the
behaviour of tidally distorted discs to be more complicated than a
simple eccentric disc, and with treatment of radiative processes the
predictions are likely to become even more complicated. Once such
models are developed further, comparisons with observation should
provide a more complete understanding of superhump phenomena.

\section{Summary}

The eclipsing novalike cataclysmic variable V348 Pup exhibits positive
superhumps. The period of these superhumps is in agreement with the
generally observed $\epsilon-P_{\rmn{orb}}$ relation for superhumpers.

Using the relation for $q$ as a function of the superhump period
excess, $\epsilon$, (Patterson 1998b) we estimate $q=0.31$. Using the
eclipse width we then estimate an orbital inclination of $i = 81\fdg 1
\pm 1\fdg 0$.

Variations in the O-C mid-eclipse times and eclipse widths strongly
suggest that V348 Pup harbours a precessing eccentric disc. We
quantify this conclusion by fitting an eccentric disc model to the
lightcurves. The relative orientation of the disc and secondary star
at superhump maximum is more easily explained by the bright spot model
for the superhump light source than the tidal heating model. A
correlation between the amplitude and orbital phase of the superhumps
is also more easily explained by the bright spot model.

Additional signals are detected at frequencies close to harmonics of
the orbital frequency. The source of these variations is currently
without conclusive explanation, but they could result from
rotationally symmetrical structure in the disc, such as spiral waves,
which have been predicted in simulations (e.g. Simpson \& Wood 1998),
and directly observed in the dwarf nova IP Peg (Steeghs, Harlaftis and
Horne 1997). They may be a result of the correlation between superhump
amplitude and orbital phase. In a high inclination system like V348
Pup the eclipse of the superhump light source may also explain these
signals.

The phasing of the superhumps in V348 Pup is like that of late
superhumps detected at the end of superoutbursts in SU UMa systems
such as OY Car and VW Hyi (Schoembs 1986 and van der Woerd et
al. 1988). The broad form of the superhump and the link between
superhump amplitude and orbital phase are also similar to late
superhumps in OY Car. By identifying the superhump mechanism in
novalikes we are likely to also understand the mechanism for late
superhumps in SU UMa systems. It is possible that this mechanism could
be different from that which produces common superhumps, generally
accepted to be a result of tidal stresses acting on an eccentric disc
(O'Donoghue 1990).

\section*{Acknowledgments}

The authors acknowledge the data analysis facilities at the Open
University provided by the OU research committee and at the University
of Sussex provided by the Starlink Project which is run by CCLRC on
behalf of PPARC. The OU computer support provided by Chris
Wigglesworth and compiling assistance from Sven Br\"{a}utigam is also
much appreciated. The {\sevensize PRIDA} eclipse mapping software
(Baptista \& Steiner 1991) was used courtesy of Raymundo Baptista. DJR
thanks Rob Hynes for being a mine of useful advice and information. We
thank Eugene Thomas and Jessica Zimmermann who carried out parts of
the observation, and the support staff at CTIO for their sterling
work. CAH thanks the Nuffield Foundation and the Leverhulme Trust for
support. DJR is supported by a PPARC studentship.


\begin{thebibliography}{99}
\bibitem{baptista} Baptista R., Steiner J. E., 1991, A\&A, 249, 284
\bibitem{baptistaet} Baptista R., Patterson J., O'Donoghue D., Buckley D., Jablonski F., Augusteijn T., Dillon W., 1996, IAUC, 6327
\bibitem{eggleton} Eggleton P.P., 1983, ApJ, 268, 368
\bibitem{hessman} Hessman F.V., Mantel K.-H., Barwig H., Schoembs R., 1992, A\&A, 263, 147
\bibitem{honey} Honey W.B., Charles P.A., Whitehurst R., Barret P.E., Smale A.P., 1988, MNRAS, 231, 1
\bibitem{horne} Horne K., 1985, MNRAS, 213, 129
\bibitem{krzeminski} Krzeminski W., Vogt N., 1985, A\&A, 144, 124
\bibitem{kuulkers} Kuulkers E., van Amerongen N., van Paradijs J., R\"ottgering H., 1991, A\&A, 252, 605
\bibitem{mk} Molnar L.A., Kobulnicky H.A., 1992, ApJ, 392, 678
\bibitem{murray96} Murray J.R., 1996, MNRAS, 279, 402
\bibitem{murray98} Murray J.R., 1998, MNRAS, 297, 323
\bibitem{murray99} Murray J.R., 1999, astro-ph/9911466
\bibitem{naylor} Naylor T., Charles P.A., Hassal B.J.M., Bath G.T., Berriman G., Warner B., Bailey J., Reinsch K., 1987, MNRAS, 229, 183
\bibitem{norton} Norton A.J., Beardmore A.P., Taylor P., 1996, MNRAS, 280, 937
\bibitem{od} O'Donoghue D., 1990, MNRAS, 246, 29
\bibitem{osaki} Osaki Y., 1996, PASP, 108, 39
\bibitem{paczynski} Paczynski B., 1977, ApJ, 216, 822
\bibitem{pat1} Patterson J., Jablonski F., Koen C., O'Donoghue D., Skillman D.R., 1995, PASP, 107, 1183
\bibitem{pat98a} Patterson J., 1998a, Disk-Instability Workshop, Kyoto, Japan
\bibitem{pat98b} Patterson J., 1998b, PASP, 110, 1132
\bibitem{patet} Patterson J., Halpern J., Shambrook A., 1993, ApJ,
419, 803
\bibitem{nr} Press W.H., Teukolsky S.A., Vetterling W.T., Flannery B.P., 1992, Numerical Recipes in C, Cambridge University Press
\bibitem{schoembs} Schoembs R., 1986, A\&A, 158, 233
\bibitem{schoembsvogt} Schoembs R., Vogt N., 1981, A\&A, 97, 185
\bibitem{snw} Simpson J.C., Wood M.A., 1998, ApJ, 506, 360
\bibitem{skillman} Skillman D.R., Harvey D.A., Patterson J., Kemp J., Jensen L., Fried R.E., Garradd G., Gunn J., van Zyl L., Kiyota S., Retter A., Vanmunster T., Warhurst P., 1998, ApJ, 503, L67
\bibitem{steeghs} Steeghs D., Harlaftis R.T., Horne K., 1997, MNRAS,
290, L28
\bibitem{tuohy} Tuohy I.R., Remillard R.J., Brissenden R.J.V., Bradt H.V., 1990, ApJ, 359, 204
\bibitem{vanderwoerd} van der Woerd H., van der Klis M., van Paradijs J., Beuermann K., Motch C., 1988, ApJ, 330, 911
\bibitem{vogt} Vogt N., 1981, ApJ, 252, 653
\bibitem{warner} Warner B., 1986, MNRAS, 219, 347
\bibitem{warnerbook} Warner B., 1995, Cataclysmic Variable Stars, Cambridge University Press, Cambridge
\bibitem{wo} Warner B., O'Donoghue D., 1988, MNRAS, 233, 705
\bibitem{whitehursta} Whitehurst R., 1988a, MNRAS, 213, 129
\bibitem{whitehurstb} Whitehurst R., 1988b, MNRAS, 232, 35

\label{lastpage}

\end{thebibliography}
\end{document}